\title{Bayesian Inference Based on Stationary Fokker--Planck Sampling}
\author{
Arturo Berrones \\
arturo.berronessn@uanl.edu.mx \\
Posgrado en Ingenier\'ia de Sistemas\\
Facultad de Ingenier\'ia Mec\'anica y El\'ectrica\\
Centro de Innovaci\'on, Investigaci\'on y Desarrollo\\
en Ingenier\'ia y Tecnolog\'ia\\
Universidad Aut\'onoma de Nuevo Le\'on\\
San Nicol\'as de los Garza, NL 66450, M\'exico
}
\date{}

\documentclass[12pt]{article}

\usepackage{graphicx,apalike}

\begin{document}
\maketitle

\begin{abstract}
A novel formalism for Bayesian learning in the context of complex inference models 
is proposed. The method is based on the use of the Stationary Fokker--Planck (SFP) approach
to sample from the posterior density. Stationary Fokker--Planck sampling generalizes
the Gibbs sampler algorithm for arbitrary and unknown conditional densities.
By the SFP procedure approximate analytical expressions for the conditionals and
marginals of the posterior can be constructed. At each stage of SFP, the
approximate conditionals are used to define a Gibbs sampling process, which is convergent to the
full joint posterior. By the analytical marginals efficient 
learning methods in the context of Artificial Neural Networks are outlined.
Off--line and incremental Bayesian inference and Maximum Likelihood Estimation from the posterior is performed
in classification and regression examples. A comparison of SFP with other Monte Carlo strategies
in the general problem of sampling from arbitrary densities is also presented. It is shown 
that SFP is able to jump large low--probabilty regions without the need of a careful
tuning of any step size parameter.
In fact, the SFP method requires only a small set of meaningful parameters which can be selected following
clear, problem--independent guidelines.  
The computation cost of SFP, measured in terms of
loss function evaluations, grows linearly with the given model's dimension.  
\end{abstract}

\section{Introduction}

Parameter inference from limited and noisy data in complex nonlinear models is a 
common and necessary step among many disciplines in modern science and engineering.
A prominent framework to extract nonlinear relations
from data is given by
Artificial Neural Networks (ANN's). These formal constructs are flexible enough to learn extremely complicated
maps. However, the potential power of ANN's is usually limited in practice because the
network size must be bounded in order to avoid poor generalization (i.e. out of sample)
performance. Authors like \cite{neal} and \cite{mackay} 
have given strong arguments that favor a Bayesian perspective, 
in which the so called {\it overtraining} problem is alleviated.
This Bayesian approach has proven its effectiveness in a number of applications \cite{auld,chib,jalobeanu}.
However, Bayesian inference based on the use of the full posterior 
density (i.e. not limited to a small subset of the posterior modes) usually 
demands very intensive computation \cite{neal}. Several techniques oriented to improve
efficiency have been proposed.
Variational Bayes \cite{beal,nakajima}, a method that has been mainly applied to particular
inference procedures like hidden Markov models and graphical models,
is a promising tool where the posterior is 
approximated by simple distributions. 
Approaches based on genetic programming also seem valuable \cite{marwala}, 
but their usefulness has not yet been established for large scale systems.
Here is introduced a new paradigm from which
a proper Bayesian estimation for large complex models can be done in 
the basis of a Gibbs sampling for the given model's weights.
This makes the procedure of relatively low computation cost and this cost increases slowly
as the inference model's dimension grows.
Moreover, from the proposed method approximate closed expressions for the 
posterior marginals can be derived. As far as the author of the present Letter knows, this
is the first approach that admits the construction of analytic expressions for the posterior
density marginals. This is useful in a number of ways. It permits the definition of efficient
maximum likelihood and incremental learning methods. Maximum likelihood can be used to drastically reduce
the computation cost for the trained inference models, making them
suitable for chip implementation, for instance. Incremental learning on the
other hand, gives a way to manage data that dynamically arrives to the inference system, enlarging
the horizons for the applications of Bayesian techniques.

\noindent The proposed method admits an arbitrary close approximation to the posterior, with 
a computational effort that is controlled through a small set of meaningful parameters.
Also, the formalism is directly connected with equilibrium statistical mechanics.
The approach is general, but in this contribution it's
validity is tested on three layered ANN's of increasing size.
A classification benchmark problem and 
two regression problems consisting of real time series with well documented
difficulty and experimental interest are considered. A discussion of the approach in the
larger context of Monte Carlo methods for sampling is also presented. 

The proposed method is based on a recently introduced algorithm
for density estimation in stochastic
search processes, namely the Stationary Fokker--Planck sampling (SFP) strategy \cite{berrones}. 
This algorithm learns the stationary
density of a general stochastic search in a potential with high dimension $V(x_1, x_2, ..., x_n, ..., x_N)$,
using only one--dimensional linear operators. 
Essentialy, SFP consists on projecting the multi--dimensional
Fokker--Planck equation associated to the stochastic search into a one--dimensional equation for the
stationary conditional cumulative distributions, $y(x_n | \{ x_{j\neq n} = 
x_j^{*} \} ) = \int _{-\infty}^{x_n} 
p(x^{'}_{n} | \{ x_{j\neq n} = x_j^{*} \} ) dx^{'}_{n}$. The starting point is the following stochastic search 
defined over 
$L_{1,n} \leq x_n \leq L_{2,n}$,
\begin{eqnarray} \label{langevin}
\dot{x}_n = - \frac{\partial V}{\partial x_n} + \varepsilon(t) ,
\end{eqnarray} 
\noindent
where $\varepsilon(t)$ is an additive noise with zero mean.
The model given by Eq.(\ref{langevin}) 
can be interpreted as an overdamped
nonlinear dynamical system composed by $N$ interacting particles. 
The temporal evolution of the probability density of such a system 
in the presence of an additive Gaussian white noise, is described 
by a linear differential equation, the Fokker -- Planck equation \cite{risken,vankampen},
\begin{eqnarray} \label{fp}
\dot{p}(x) = \sum_{n=1}^N \frac{\partial}{\partial x_n} \Big[
\frac{\partial V}{\partial x_n} p(x) \Big] + D \sum_{n=1}^N \sum_{m=1}^N
\frac{\partial^2 p(x) }{\partial x_n \partial x_m },
\end{eqnarray}
\noindent
where $D$ is a constant, called diffusion constant, that is proportional to the noise
strength. The direct use of Eq. (\ref{fp})
for optimization or deviate generation purposes would imply 
the calculation of high dimensional integrals.
It results numerically much less demanding to perform the 
following one dimensional projection of Eq. (\ref{fp}).
Under very general conditions (e. g., the absence of infinite 
cost values), the equation (\ref{fp}) has a stationary solution
over a search space with reflecting boundaries \cite{risken,grasman}. 
The stationary conditional probability density
satisfy the one dimensional Fokker -- Planck equation
\begin{eqnarray}\label{sfp}
D\frac{\partial p(x_n | \{ x_{j\neq n} = x_j^{*} \} )}{\partial x_n}
+p(x_n | \{ x_{j\neq n} = x_j^{*} \} ) \frac{\partial V}{\partial x_n} = 0 .
\end{eqnarray}
\noindent
From Eq. (\ref{sfp}) follows 
a linear second order 
differential equation for the cumulative distribution $y(x_n | \{ x_{j\neq n} = 
x_j^{*} \} ) = \int _{-\infty}^{x_n} 
p(x^{'}_{n} | \{ x_{j\neq n} = x_j^{*} \} ) dx^{'}_{n}$,
\begin{eqnarray}\label{sfpm}
\frac{d^{2}y}{dx_{n}^{2}}+\frac{1}{D}\frac{\partial V}{\partial x_n}\frac{dy}{dx_n}
= 0 ,\\ \nonumber
\\ \nonumber
y(L_{1,n})=0, \quad y(L_{2,n})=1 .
\end{eqnarray}
\noindent
Random deviates can be drawn from the density $p(x_n | \{ x_{j\neq n} = x_j^{*} \} )$ by the fact 
that $y$ is a uniformly distributed random variable
in the interval $y \in [0, 1]$. Viewed as a function of the random variable $x_n$, 
$y(x_n | \{ x_{j\neq n} \})$ can be approximated through a 
linear combination of functions from a complete set 
that satisfy the boundary conditions in the interval of interest,
\begin{eqnarray}\label{set}
\hat{y}(x_n | \{ x_{j\neq n} \})=\sum_{l=1}^{L} a_l \varphi _l  ( x_n  ) .
\end{eqnarray}
\noindent
Choosing for instance, a basis in which $\varphi _ l ( 0 ) = 0$, the $L$ coefficients
are uniquely defined by the evaluation of Eq. (\ref{sfpm}) in $L-1$ interior points. In this way, the 
approximation of $y$ is performed by solving a set of $L$ linear algebraic equations, involving
$L-1$ evaluations of the derivative of $V$.

The SFP sampling is based on the iteration of the following steps:

\noindent
{\bf 1)} Fix the variables $x_{j\neq n} = x_j^{*}$ and approximate $y(x_n | \{ x_{j\neq n} \})$
 by the use of formulas
(\ref{sfpm}) and (\ref{set}).

\noindent
{\bf 2)} By the use of $\hat{y}(x_n | \{ x_{j\neq n} \})$ 
construct a lookup table in order to
generate a deviate 
$x_n^{*}$ drawn from the stationary distribution
$p(x_n | \{ x_{j\neq n} = x_j^{*} \})$.

\noindent
{\bf 3)} Update $x_n = x_n^{*}$ and repeat the procedure for a new variable $x_{j \neq n}$.

\noindent

The fundamental parameters of SFP sampling, $L$ and $D$, 
have a clear meaning, which is very helpful for their selection. The diffusion 
constant ``smooth'' the density. This is evident by taking the limit $D \to \infty$ in
Eq. (\ref{sfpm}), which imply a uniform density in the domain. The number of base functions
$L$, on the other hand, defines the algorithm's capability to ``learn'' more or less
complicated density structures. Therefore, for a given $D$, 
the number $L$ should be at least large enough to assure 
that the estimation algorithm will generate valid distributions $y(x_n | \{ x_{j\neq n} \})$. 
A valid distribution
should be a monotone increasing continuos function that satisfy the boundary conditions.
The parameter $L$ ultimately determines the computational cost of the procedure, because 
at each iteration a system of size $\propto L$ of linear algebraic equations must be solved
$N$ times.
Therefore, the user is able to control the computational cost through the interplay of the
two basic parameters: for a larger $D$ a smoother density should be estimated, 
so a lesser $L$ can be used.

It should be 
noticed that SFP is a generalization of Gibbs sampling. Therefore, the deviates generated by the
iterative procedure are in the long run sampled from the full joint equilibrium density
\begin{eqnarray}\label{boltzmann}
p(x_1, x_2, ..., x_n, ..., x_N) = \left( \frac{1}{Z} \right) \exp(-V / D),
\end{eqnarray}
\noindent where $Z$ is a normalization factor.
Additionaly,
a convergent representation for $y(x_n)$ is obtained after taking the average
of the coefficients $a$'s in the expansion (\ref{set})
over the iterations \cite{berrones}, 
\begin{eqnarray}\label{setav}
\left< \hat{y} \right> =\sum_{l=1}^{L} \left< a_l \right> \varphi _l  ( x_n  ) 
\to y(x_n) ,
\end{eqnarray}
Under general conditions,
the Gibbs sampler converges at a geometric rate \cite{roberts,canty} and there is some evidence that 
this fast convergence is shared by SFP \cite{berrones2}. In the next section the properties
of the SFP sampler are studied in the wider context of Monte Carlo methods. Thereafter the general toolkit for
Bayesian inference based on SFP is developed and tested.

\section{SFP sampler}

\begin{figure}
	\centering
		\vskip 0.7cm
		\includegraphics[width=.7\textwidth]{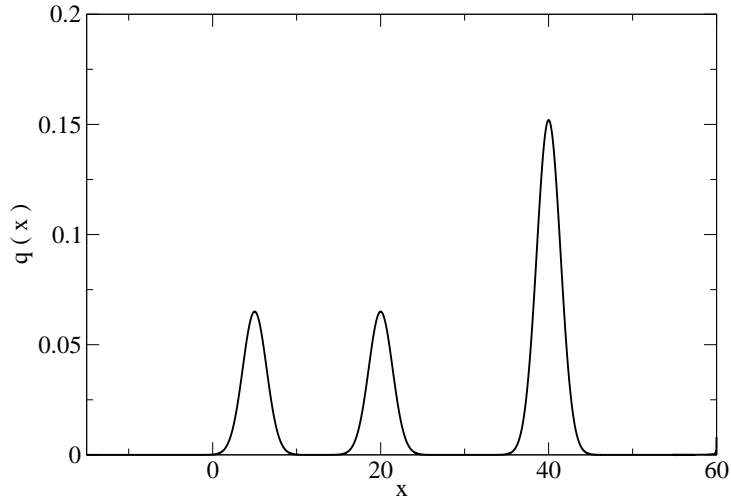}
	\caption{Density estimated by SFP for the mixture density problem.}
	\label{density1}
\end{figure}

\begin{figure}
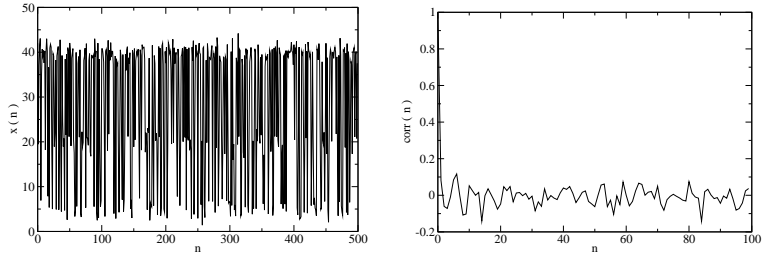

	\centering
		\vskip 0.7cm
		\includegraphics[width=.35\textwidth]{sample1.eps}
		\hskip 0.3cm
		\includegraphics[width=.35\textwidth]{correlation1.eps}
	\caption{Sample generated by SFP and it's autocorrelation function for the mixture
		density problem.}
	\label{sample1}
\end{figure}

\begin{figure}
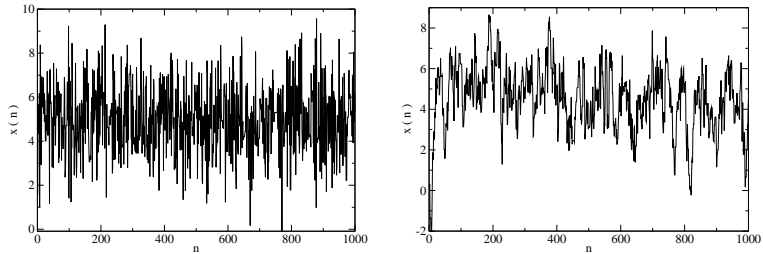

	\centering
		\vskip 0.7cm
		\includegraphics[width=.35\textwidth]{metrop.eps}
		\hskip 0.3cm
		\includegraphics[width=.35\textwidth]{hmc.eps}
	\caption{Samples generated by Metropolis (first graph) and HMCMC (second graph) for the 
		mixture density problem. Both samplings remain trapped in a local region of the relevant search space.}
	\label{mc}
\end{figure}

In order to sample a given distribution $\pi (\vec{x})$, the potential function that enters into SFP should
be given by
\begin{eqnarray}\label{sampler1}
V(\vec{x}) = - \ln \pi (\vec{x}), 
\end{eqnarray}
\noindent using $D = 1$. The correct normalization is directly obtained by construction without explicitly calculating it nor
including it in the definition of $V$. 
The advantages of SFP sampling with respect to previous Monte Carlo methods are first illustrated
through an important class of probability densities, namely the separable probability densities of the form
\begin{eqnarray}\label{separable}
\pi (\vec{x}) = \prod _{i=1}^{N} f(x_i).
\end{eqnarray}
\noindent In this case SFP converges in a single iteration. This result follows from the fact that the dynamics 
(\ref{langevin}) associated to the random search decouples into $N$ independent one dimensional stochastic
differential equations, which makes the
Gibbs sampling stage (steps 2 and 3) of SFP unnecessary.
Because $D$ is fixed, SFP requiers a single parameter, $L$. But $L$ simply refers to the number of base functions
used in expansion (\ref{set}), so in SFP there is no need to adjust a step size parameter. Step
size parameters are difficult to tune because their correct selection depends on the actual variance of the
sampled distribution, which in general is unknown. The parameter $L$ on the other hand, is selected by
objective criteria that are independent from any knowledge about the sampled density. Specifically, 
$L$ should be large enough such that the observed marginals are valid probability densities.
Consider the next example, a one dimensional mixture density
given by 
\begin{eqnarray}\label{mixture}
q(x) = \sum _{i=1}^{R} r_i f_i(x),
\end{eqnarray}
\noindent where $f_i(x)$ are normal densities $N(\mu _i, \sigma _i)$. It has been pointed out in the literature that mixture 
densities with well separated modes are difficult to sample by Monte Carlo methods \cite{celeux,marin}. 
Consider a case
with three modes, which result from the mixture of normal densities 
$ \pi(x) = 0.2 N(5, 2) + 0.2 N(20, 2) + 0.6 N(40, 2)$.
This mixture density has mean and standard deviation $\mu = 29$ and
$\sigma = 14.35$ respectively.
The Figure \ref{density1} shows the resulting estimation of the density after a single SFP iteration with $L = 1100$.
In all the numerical experiments discussed in this Letter,
a Fourier basis is used in formula (\ref{set}) to approximate the distributions produced by SFP sampling,
\begin{eqnarray}\label{set2}
\hat{y}=\sum_{l=1}^{L} a_l \sin \left ( 
(2l-1)\frac{\pi (x_n - L_{1,n} )}{2(L_{2,n}-L_{1,n})}
\right ).
\end{eqnarray}
In the estimation of the mixture density,
the parameter $L$ has been chosen by growing it's size by $200$ units per experiment, starting at $L = 100$. 
After each experiment the resulting density is visually inspected and it's variance is calculated by 
direct integration from the formula Eq. (\ref{set}). It's selected the
minimum value of $L$ which produce a valid density with positive variance. Each SFP iteration takes around
$4$ seconds in the equipment used (details of which are given in Section \ref{experiments}).
Points from
this density are easly drawn by the construction of a lookup table for the cumulative distribution Eq. (\ref{set})
to generate each deviate as in step 2 of SFP.
Figure \ref{sample1} shows a sample of $500$ points whose generation took a fraction of a second.
The auto--correlation 
function for the sample is also plotted. No significant correlations 
appear between the points in the sample.
The average and standard deviation estimated from the sample are $\hat{\mu} = 27.31$ and
$\hat{\sigma} = 14.6$ respectively.
The same mixture density sampling problem is studied by the Metropolis and Hybrid Markov--Chain Montecarlo (HMCMC)
algorithms. 
The implementations provided in the
``Software for flexible Bayesian modeling and Markov chain sampling'' developed by Radford Neal
(downloadable at http://www.cs.toronto.edu/$\sim$radford/) are used. The parameters for the Metropolis
and HMCMC have been carefully tuned in order to have acceptable rejections rates, which are $0.7$ for
Metropolis and $0.78$ for HMCMC. Samples of $1000$ points provided by both methods are presented in Fig. \ref{mc}. 
The estimated average from both samples is around the value of $5$.
These results show one of the main drawbacks in many Monte Carlo strategies, namely the possibility to reach a state
of apparent equilibrium which in fact is unrepresentative of the whole density.
The mixture density example illustrates what might be one of the most
promising features of SFP: it's capacity to jump large regions of low probability, reducing the danger of
getting trapped into local high probability regions. 

For multidimensional non--separable distributions the Gibbs sampling stage of SFP is essential. 
Consider the following
example, provided in the tutorial of the ``Software for flexible Bayesian modeling and Markov chain sampling'',
\begin{eqnarray}\label{ring}
V(\vec{x}) = \frac{1}{2}(x^2 + y^2 + z^2) + (x+y+z)^2 + 10000/(1+x^2+y^2+z^2), 
\end{eqnarray}
which gives a three--dimensional ring density. The Table \ref{table-ring} lists the results of
the estimated expectations for each one
of the variables from $2000$ points generated by the samplers without rejecting initial points. 
By symmetry, the correct expected values are equal to zero. For SFP $L = 150$ in the search space defined by
the cube $[-20, 20]^3$. The parameters for HMCMC and Metropolis are the same as discussed in the software's tutorial.

\begin{table}
\begin{center}

\begin{tabular}{|c|c|c|c|c|c|c|c|c|c|c|}  \hline
&	\textbf{E(x)} &  \textbf{E(y)} &   \textbf{E(z)}         	\\
	\hline 
\textbf{SFP} &	-0.04 &  -0.5 &  0.54	\\
	\hline 
\textbf{HMCMC} & 0.48  & -0.18  &  -0.31	\\
	\hline 
\textbf{Metropolis} & 1.92 &  -0.61 & -1.35 	\\
	\hline 
\end{tabular} 
\vspace{10pt}
\caption{\label{table-ring} Estimations of the mean values by SFP, HMCMC and Metropolis algorithms
	in the three--dimensional ring distribution problem. By the symmetries of the sampled 
	distribution, the true mean values are equal to zero for all the three variables. SFP and
	HMCMC display similar results.} 
\end{center}
\end{table}  

\begin{figure}
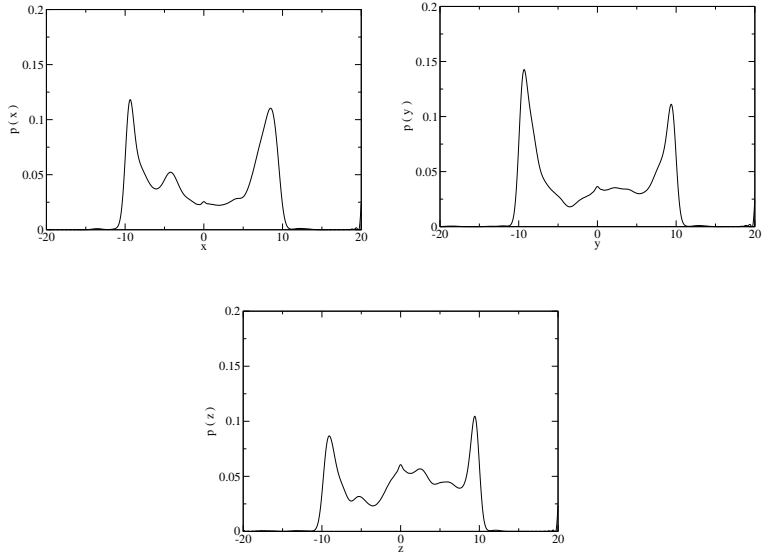

	\centering
		\vskip 0.7cm
		\includegraphics[width=.35\textwidth]{ringx.eps}
		\hskip 0.3cm
		\includegraphics[width=.35\textwidth]{ringy.eps}
		\vskip 0.7cm
		\includegraphics[width=.35\textwidth]{ringz.eps}
	\caption{Marginals estimated by SFP for the three--dimensional ring problem.}
	\label{marginals-ring}
\end{figure}

\begin{figure}
	\centering
		\vskip 0.7cm
		\includegraphics[width=.7\textwidth]{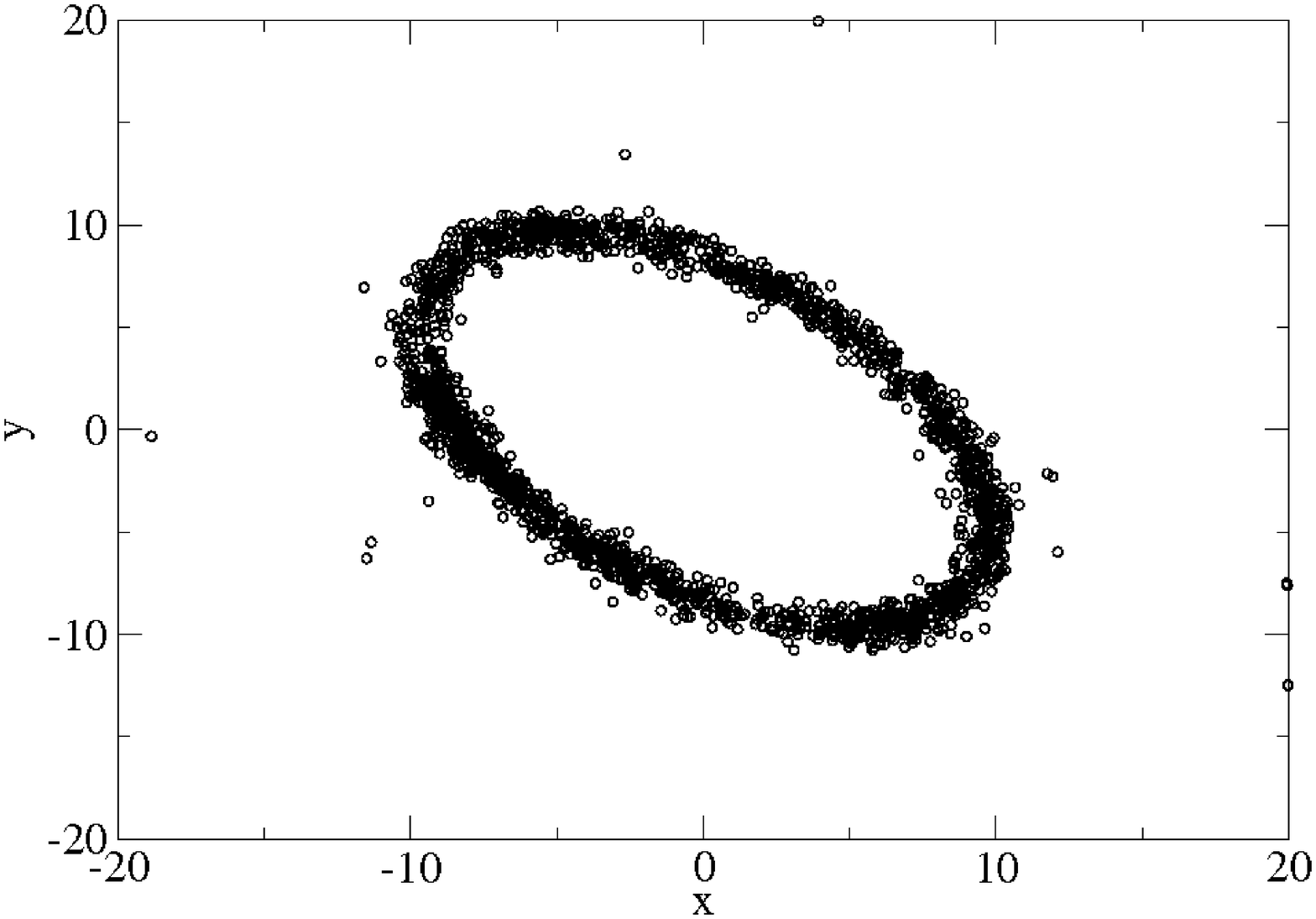}
	\caption{Plot in the $x-y$ plane of a sample generated by SFP of the three--dimensional ring problem.}
	\label{x-y}
\end{figure}

The SFP and HMCMC methods display similar estimations of the expected values while the Metropolis
algorithm shows a significantly poorer performance. The power spectrum of the samples generated by the
different methods has been studied. It has turned out that SFP has a power spectrum which is consistent
with a random walk, a behavior shared with Metropolis and with many other Monte--Carlo approaches. The HMCMC method
on the other hand, shows a power spectrum which indicates exponential decay in the autocorrelations. However,
perfectly independent samples of the marginals (which might be of interest in several applications) 
can be generated by the use of SFP. The estimated analytic forms of the marginal densities of the ring distribution are 
shown in Fig. \ref{marginals-ring}. Figure \ref{x-y} 
is a plot of the sample generated by SFP in the $x-y$ plane, which indicates that SFP captures the
interactions among variables.

\begin{figure}
	\centering
		\vskip 0.7cm
		\includegraphics[width=.7\textwidth]{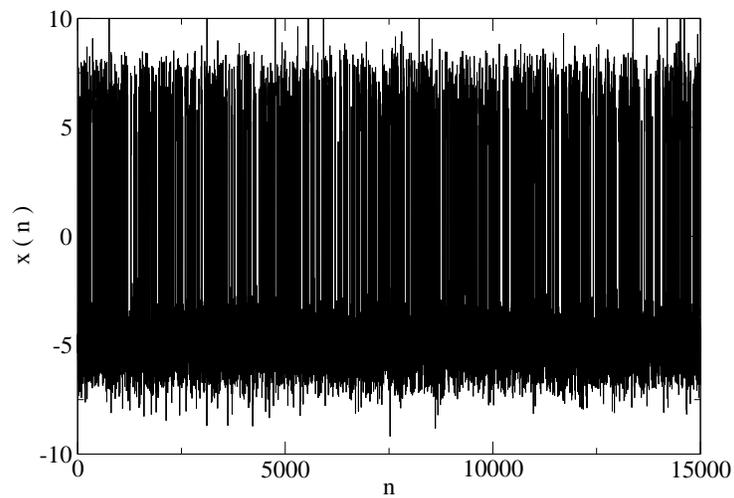}
	\caption{Sample generated by SFP for a mixture of two bivariate Gaussians.}
	\label{bivariate1}
\end{figure}

\begin{figure}
	\centering
		\vskip 0.7cm
		\includegraphics[width=.7\textwidth]{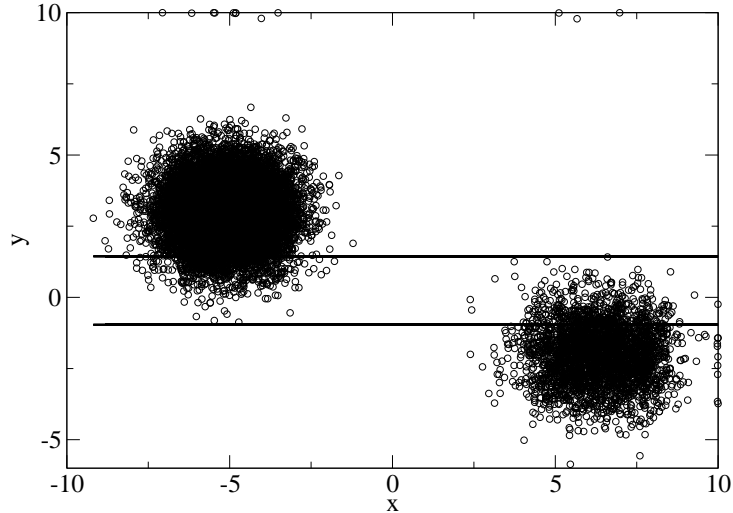}
	\caption{Plot in the $x-y$ plane of a sample generated by SFP for a mixture of two bivariate Gaussians.
		The straight lines show that there is a positive probability that 
 		during the iterations of a given SFP run a conditional $p(x | y)$ that detect
		both modes will be drawn.}
	\label{bivariate2}
\end{figure}

\begin{figure}
	\centering
		\vskip 0.7cm
		\includegraphics[width=.7\textwidth]{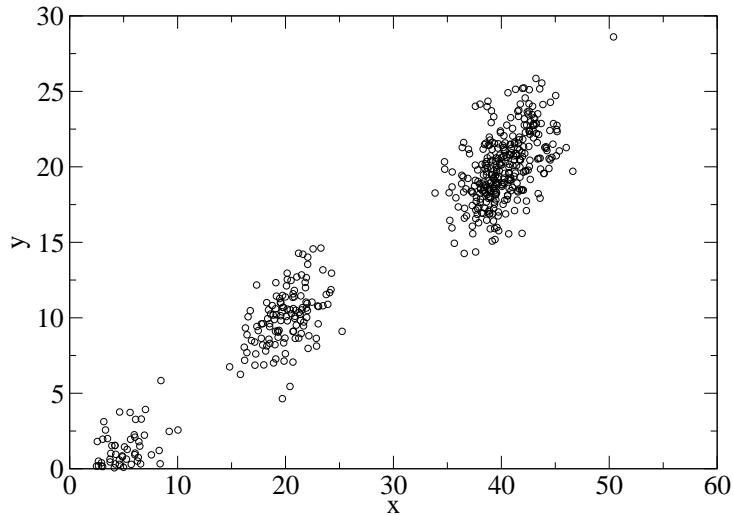}
	\caption{Plot in the $x-y$ plane of a sample generated by SFP for a mixture of three bivariate Gaussians.}
	\label{bivariate3}
\end{figure}

An additional comparison of SFP with the family of Monte Carlo strategies based on Langevin
diffusions is given in the ground of
the sampling of mixture densities in several dimensions. As already pointed out, this is a difficult task for
many samplers, specially when the mixture has well separated modes. In \cite{roberts2} is proposed a 
mixture of two bivariate Gaussians in order to test the performance of different  
Langevin-diffusion based algorithms. The mixture is the following,
$\pi (\vec{x}) = 0.5 N (\vec{\mu}_1, \Sigma) + 0.5 N (\vec{\mu}_2, \Sigma)$ with 
$\vec{\mu}_1 = (6, -5)^{T}$, $\vec{\mu}_2 = (-2, 3)^{T}$ and $\Sigma = \bar{I}$. The Figure \ref{bivariate1} shows
the generation of $15000$ sample points by SFP with $L = 200$ in a search space $[-10, 10]^2$. 
This graph can be directly compared with the
experiments reported in \cite{roberts2}. It is clear that SFP is able to very quickly find the two modes of the 
mixture. The rapid switching between modes seems to outperform Langevin diffusions. 
In the Figure \ref{bivariate2} the sample generated by SFP in the $x-y$ plane is plotted. 
The straight lines show that there is a positive probability that 
during the SFP iterations a conditional $p(x | y)$ that detect
both modes is drawn. This is how SFP eventually mix over the two modes. In this sense, SFP connects
the search space through the $x$ direction given that there is some overlap of the modes in the $y$ direction.
Some deviates lie on the 
border of the search space. This effect is a consequence of the discretization used to solve the 
stationary Fokker-Planck equation.
With a larger $L$, the marginals are approximated more closely and the border effect can be corrected, at the 
cost of a larger computation effort.
A very difficult task occurs when a
mixture effectively split a large search space into several distant unconnected regions.
The following case is considered,
$\pi (\vec{x}) = 0.2 N (\vec{\mu}_1, \Sigma) + 0.2 N (\vec{\mu}_2, \Sigma) + 0.6 N (\vec{\mu}_3, \Sigma)$ with 
$\vec{\mu}_1 = (5, 0)^{T}$, $\vec{\mu}_2 = (20, 10)^{T}$, $\vec{\mu}_3 = (40, 20)^{T}$ and
\begin{eqnarray}
\Sigma = \left( \begin{array}{ccc}
2 & 0.5  \\
0.5 & 2 \\
\end{array} \right)
\end{eqnarray}
\noindent For this example SFP is unable to find the three modes in a reasonable amount of time. It is however possible to
easily sample this density if values $D > 1$ are allowed. By exploring with diffusion coefficients slightly larger than
$D=1$ the user is able to consider approximated densities with extra noise in which the search space is connected.
In the Figure \ref{bivariate3} a $500$ points sample with $L = 300$ and $D = 1.1$ 
in the search space $[0,60]^2$ is plotted, showing that SFP is capable to detect all of the modes,
with the sample points distributed among them in consistency with the underlying mixture.
A more careful comparison of SFP with several Monte Carlo methods
for sampling is a topic which deserves further research.
From the results given so far it is however possible to establish a list of general statements about
how SFP relates with other Monte Carlo procedures:

\begin{itemize}
\item SFP generalizes the Gibbs sampler for cases in which the conditionals are not explicitly known. Therefore SFP inherits many of the
advantages and drawbacks of the Gibbs sampler. 

\item There is no need for a step size parameter, which is one of the main
computational advantages of the Gibbs sampler and SFP.

\item Like for the Gibbs sampler, the samples generated by SFP display a random walk behavior. In this regard, the Monte Carlo methods
which include momentum appear to be superior than SFP when almost independent samples are required. However, perfectly 
independent samples
for the marginals can be generated by SFP (a property not shared by the Gibbs sampler).

\item Like the Monte Carlo methods based on Langevin diffusions, SFP requires the numerical evaluation of the gradient of the sampled density.

\item SFP involves the solution of $\propto LN$ linear algebraic equations (where $N$ is the problem's dimension and $L$ is the
number of basis functions used in order to approximate the solution of a one dimensional stationary
Fokker--Planck equation) 
for the generation of each sample point. This is a sophisticated operation 
if compared to what is required by most of Monte Carlo techniques. The Metropolis-Hastings type algorithms are particularly
much more simple to implement than SFP.
\end{itemize}

It is of particular interest to more deeply investigate the possible applications of SFP in mixture models, which
offer considerable challanges to Monte Carlo techniques despite being
among the most widely used statistical methods for the study of complex systems \cite{marin}.
It appears to the author of the present Letter that this line of thought would be 
worth to be explored in the future.
The next sections discuss the application of SFP in large dimensional problems in the context of Bayesian inference for neural networks, 
which is the main topic of the present Letter.

\section{Bayesian inference in complex models}

In the Bayesian approach to learning SFP enters naturally by the use of a potential
$V$ such that the Boltzmann--type density  (\ref{boltzmann}) reproduce the posterior of interest.
Consider the following setup. 
The relation between a vector of inputs $\vec{x}$ and a vector of outputs $\vec{y}$ is characterized by 
the function $F(\vec{x},\vec{w})$ where $\vec{w}$ is a set of parameters.
A suitable learning task is the estimation of the predictor
$E[F(\vec{x})]$. In the Bayesian framework, these type of estimations are carried out by the update 
of prior probabilities 
considering the empirical evidence at hand. Let $A$ be a set of empirical observations. 
The probability density of the function parameters $\vec{w}$ given $A$ is written as
\begin{eqnarray}\label{bayes1}
p(\vec{w} | A) = p(\vec{w}) p(A | \vec{w})
\end{eqnarray}
\noindent It follows from Eq. (\ref{boltzmann}) that
the potential function that enters into SFP is given by
\begin{eqnarray}\label{bayes2}
V(\vec{w}) = - [\ln p(\vec{w}) + \ln p(A | \vec{w})].
\end{eqnarray}
\noindent In the SFP framework proposed here, three schemata naturally arise for the learning of complex
functions from data, which are discussed below.

\subsection{Bayesian inference by sampling from the posterior}

The SFP iterations converge to the sampling of the posterior, which follows from the fact that SFP is a particular
form of the Gibbs algorithm, for which this property holds in general \cite{roberts,canty}. 
Moreover, it has been rigorously demonstrated that 
under general conditions, Gibbs sampling converges
at a geometric rate \cite{roberts,canty}. 
In this way, the deviates generated by $\tau$ SFP iterations (perhaps 
after rejecting a number of the initial
set of iterations) can be used in order to estimate the integral
\begin{eqnarray}\label{optimalY}
\langle \vec{y} \rangle = \int  F(\vec{x},\vec{w}) p(\vec{w} | S) d\vec{w},
\end{eqnarray}
\noindent by the average
\begin{eqnarray}\label{optimalY2}
\langle \vec{y} \rangle \approx \frac{1}{\tau} \sum_{t=1}^{\tau}  F(\vec{x},\vec{w}_t) .
\end{eqnarray}  

\subsection{Maximum Likelihood Estimation (MLE) from the posterior marginals}

The SFP framework admits the construction of explicit expressions for the marginals of the posterior, 
given by equation (\ref{setav}).
These marginals can easily be maximized by one dimensional optimization methods to provide a MLE
for each of the weights. Moreover, the weigth moments give Bayesian corrections to the MLE, in terms
of the expansion
\begin{eqnarray}\label{smallFluctuation}
\langle F(\vec{x}) \rangle \approx F(\vec{x},\vec{w_o}) + 
\nabla F(\vec{x},\vec{w_o}) (\langle \vec{w} \rangle - \vec{w_o}) \\ \nonumber
+ \frac{1}{2} (\langle \vec{w} - \vec{w_o} \rangle)^{T} \nabla ^{2} 
F(\vec{x},\vec{w_o}) (\langle \vec{w} - \vec{w_o} \rangle) + ...
\end{eqnarray}
\noindent If the weights are assumed to be independent, all the 
statistical moments involved in the expansion (\ref{smallFluctuation})
can be calculated by solving the corresponding one dimensional integrals that follow from the
marginals (\ref{setav}). If weight independence is not assumed, the cross moments can be estimated
through averages similar to (\ref{optimalY2}). For instance, covariances are
estimated by
\begin{eqnarray}\label{wcov}
\langle w_n w_m \rangle - \langle w_n \rangle \langle w_m \rangle 
\approx 
\frac{1}{\tau} \sum_{t=1}^{\tau} w_n w_m - \langle w_n \rangle \langle w_m \rangle .
\end{eqnarray}
\noindent The result of MLE 
is a formula by which there is no need to perform additional numerical averages 
in order to evaluate the trained network at a given input. Therefore MLE may be useful in applications
in which the trained model should run under limited computation resources, like 
for instance in embbeded systems.

\subsection{Incremental Bayesian inference}

To have explicit expressions for the posterior's marginals turns out to be advantageous in several respects,
as already pointed out for MLE.
In particular, the marginals can be viewed as a way to encode the characteristics learned by the inference model
from the given sample, through the average coefficients $\left < a \right >$'s. 
Consider a sample with $A$ observations from which a number of
SFP iterations have been runned. 
If a new observation arrives, the weigths should be updated according to Bayes theorem. The
marginals learned by SFP provide quite a natural way to define the necessary priors. The following mechanism is
proposed. From Eq. (\ref{bayes2}) the potential function can be written,
\begin{eqnarray}
V(\vec{w} | A) = - [\ln p(\vec{w}) + \ln p(A | \vec{w}) ].
\end{eqnarray}
\noindent As a result of running SFP with this potential, expressions for the marginals of
the weight's densities $p(w_i | A)$ are obtained by derivation from the estimated marginal distributions (\ref{setav}).
When a new observation arrives, it's proposed to construct a new prior with these learned marginals in
the following way,
\begin{eqnarray}
p(\vec{w}) = \prod _{i=1}^{N} p(w_i | A).
\end{eqnarray}
Therefore for the sample with $A+1$ observations, 
the potential is updated like,
\begin{eqnarray}\label{bayesUpdate}
V(\vec{w} | A+1) = 
- \left [ \sum _{i=1}^{N} \ln p(w_i | A) + \ln p(A+1 | \vec{w}) \right ],
\end{eqnarray}
\noindent Incremental learning is useful in a context in which the sample is not completely known
before the training begins, but data arrives in a dynamic fashion. 
In such settings it is necessary to update an existing inference model under the presence of the
new data.
Clearly the procedure presented here is able to handle as many previous data as desired, depending in
the sample chosen to enter into the likelihood $p(data | \vec{w})$. In the numerical experiments considered in
this Letter, only the incremental version in which the likelihood depends on the entire accumulated sample
is considered. Other possibilities, like for instance the on--line version, in which
only the last datum enters into the likelihood, appear to be relevant in the context of 
strongly unstationary systems.
These aspects are expected to be investigated in a future work.

\section{Experiments}\label{experiments}

In the numerical experiments three--layered ANN's with hyperbolic tangent activation functions are considered, except on the
output layer, where linear activation functions are used for regression and soft--max activation
functions are employed in classification.
Quadratic loss has been used,
\begin{eqnarray}
- \ln p(A | \vec{w}) = \frac{1}{D} \frac{1}{A} \sum_{a=1}^{A} \parallel \vec{y}_a - F(\vec{x}_a, \vec{w}) \parallel ^{2}.
\end{eqnarray}
\noindent This choice corresponds to a Gaussian likelihood with an error's variance of $\sigma^2 = D / 2$.
Notice that the parameter $D$ only affects the likelihood part of the posterior.
Parameter selection for the learning of an appropriate posterior density depends on the desired computational
effort per iteration of SFP, determined by $L$ and on the statistical 
properties of the resulting posterior, which are controlled by $D$.
In the following experiments, these parameters have been selected by running a single SFP iteration
and then observing the learned posterior marginal densities. Keeping $L$ fixed, $D$ is diminished and chosen to be
the lesser $D$ for which the resulting marginal is a valid probability density.
The initial prior densities for the weights are given by uniform densities 
in the interval $[-1, 1]$. This is at some extent an arbitrary and uninformative choice, 
because no distinction
is made between the different types of weights (e. g. biases or input connections).
The first $15$ iterations of SFP are rejected for the sampling from the posterior method.
In the case of incremental Bayesian inference the weights are updated after each SFP iteration, performing
a single SFP iteration per sample, where each new sample is given by the previous one plus a single 
new datum. 
For MLE the first order of the expansion (\ref{smallFluctuation}) 
is used for network evaluation purposes.
The experiments have been carried on a
standard PC with a 3 Ghz processor and 512 MB of RAM memory, running Linux. The SFP method and the necessary
classes for three--layered ANN's have been programmed in the Java language. The Colt numerical library
(http://acs.lbl.gov/$\sim$hoschek/colt/) has been used for the solution of linear systems and random 
number generation required by SFP.  

The first example consists on a well known classification task that has already been used to test
Bayesian approaches to learning.
Two difficult signal prediction tasks are also considered. A common problem in nonlinear signal analysis is
the forecast of short and noisy time series \cite{kantz}. 
Data of this kind play a central role in scientific, medical and
engineering applications \cite{kantz}. 
A central difficulty that arises in this context comes from the fact that nonlinear dynamical
systems may exhibit deterministic behavior that is statistically equivalent to noise. In short and noisy 
samples this behavior can easily mislead a given predictive model, giving rise to strong overfitting.

\subsection{Classification on the forensic glass data}

The ``Glass Identification'' data set (downloadable at the 
UC Irvine Machine Learning Repository, http://archive.ics.uci.edu/ml/), 
consists on $214$ instances of glass fragments found at the scene of
a crime. The task is to identify the origin of each fragment based on refractive index and chemical composition.
Reliable identification can be valuable as evidence in a given criminological investigation.
This data set has been used to test several nonlinear classifiers, including Bayesian approaches \cite{neal,ripley}.
In accordance to previous authors, the following classes have been considered: float--processed window glass, 
non--float--processed window glass, vehicle glass, and other. The headlamp glass has been discarded, leaving a total
of $185$ instances. 
The 
attributes are the refractive index and the percent by weight of oxides of Na, Mg, Al, Si, K, Ca, Ba and Fe. 
These values have been normalized to have zero mean and unit variance. The classifiers studied here consist
on three layered ANN's with soft--max activation functions for the output layer. The performance is measured on terms
of the fraction of mis--classification, where the
attribute with the largest soft--max value is interpreted as the output of the ANN.
In each experiment,
the data has been splitted on training and test sets in the following manner:
Float--processed window glass: $30$ train, $40$ test. 
Non--float--processed window glass: $39$ train, $37$ test.
Vehicle glass: $9$ train, $8$ test.
Other: $11$ train, $11$ test.
Ten experiments for a network with $6$ hidden layers
have been performed and the resulting mis--classification's average and variablity are reported on
Table \ref{table-glass}. The compared methods are Maximum Likelihood Estimation based in the SFP posterior marginals (SFP-MLE),
SFP sampling from the full posterior (SFP-S), incremental SFP learning (SFP-I) and two versions of Hybrid Markov--Chain
Montecarlo (HMCMC1 and HMCMC2). 
An implementation of HMCMC provided by the original author of the method in his
``Software for flexible Bayesian modeling and Markov chain sampling'' is used.
The two HMCM methods differ in the their selected parameters. HMCMC1 follows
the description given in \cite{neal} for the ``Glass Identification'' data set using
non vague priors. 
According to the author \cite{neal} the chosen parameters are such that
the HMCMC procedure converges to the correct posterior distribution with a very high degree of confidence. 
In HMCMC2 the parameters provided by the the author for the classification example used 
in the description of his software are used.
These parameters imply a lesser number of HMCMC iterations and shorter computation times at the cost of a
higher risk of inadequate convergence.  

\noindent The SFP parameters for this experiment are: $L = 100$, $D =  5e-4$ with a number of SFP iterations
$M = 100$.
For comparision purposes, some previously published results \cite{neal,ripley}
over a single experiment using
other approaches are included. SFP-S and HMCMC show the best performance, but HMCMC1 takes $55.6$ minutes
in order to complete training for each experiment, 
while SFP-S took $5$ minutes. For this example it appears however that
the faster version HMCMC2 adequately converges to the posterior of interest, showing similarly good 
performance. HMCMC2 took $4.3$ minutes of computation time.

\begin{table}
\begin{center}

\begin{tabular}{|c|c|c|c|c|c|c|c|c|c|c|}  \hline
&	\textbf{mis--classification rate}	\\
	\hline 
\textbf{SFP-S} &	0.32 $\pm$ 0.04   	\\
	\hline 
\textbf{SFP-MLE}&	0.33 $\pm$ 0.03   	\\
	\hline 
\textbf{SFP-I} &	0.33 $\pm$ 0.03  	\\
	\hline 
\textbf{HMCMC1} &	0.32 $\pm$ 0.04  	\\
	\hline 
\textbf{HMCMC2} &	0.32 $\pm$ 0.04  	\\
	\hline 
\textbf{From base rates in training set}&	  0.61 	\\
\hline
\textbf{Max. penalized likelihood ANN, 6 hidden units}&	 0.38 	\\
\hline
\end{tabular} 
\vspace{10pt}
\caption{\label{table-glass} Performance of SFP Bayesian learning in the 
classification of the forensic glass data. The reported mis--classification rate for SFP
and HMCMC
is based on $10$ independent experiments. For comparision purposes, 
previously published results over a single experiment with other methods are included.} 
\end{center}
\end{table}

\subsection{Prediction of human breath rate}

The human breath rate signal is part 
of a well known multichannel physiological data set provided for the Santa Fe Institute
time series competition in 1991/92 \cite{weigend}. 
The data set contains the instantaneous heart rate, air flow and blood
oxygen concentration, recorded twice a second for one night from a patient that shows sleep apnea 
(periods during which he takes a few quick breaths and then stops breathing for up to 45 seconds).
The experimental system is clearly non--stationary. Following \cite{kantz} 
here has been used an approximately stationary
sample of the air flow through the nose of the human subject. Starting at measurement 12750, 1000 data points
have been selected. The first 500 points are used as a training set. The data has been normalized such that
it has unit variance and zero mean.

\noindent There is a large amount of evidence indicating
that this multichannel physiological data contains nonlinear structure and a strong random component \cite{kantz}. 
These
facts are confirmed in our sample
(see Table \ref{table-breath}).

\begin{table}
\begin{center}

\begin{tabular}{|c|c|c|c|c|c|c|c|c|c|c|}  \hline
&	\textbf{In sample MSE}&	\textbf{Predicted MSE}&	\textbf{Out of sample MSE}		\\
	\hline
\textbf{SFP-S} &	 --   &	0.0550&	0.0708	\\
	\hline 
\textbf{SFP-MLE}&	 --   &	0.1000&	0.2001	\\
	\hline 
\textbf{SFP-I} &	 --   &	0.0804&	0.0946	\\
	\hline 
\textbf{HMCMC} &	 --   &	0.2511&	0.2379	\\
	\hline 
\textbf{BFGS}  &	 0.0400   &	--    &	0.4100	\\
	\hline 
\textbf{AR(16)}&  --  &	0.2000 &	0.3205	\\
	\hline 
\textbf{LNP}&	--    &	--    &	 0.2400       \\
	\hline 
\end{tabular} 
\vspace{10pt}
\caption{\label{table-breath} Performance of SFP and HMCMC Bayesian learning in comparision with a standard
training procedure in an ANN with $10$ inputs, $20$ hidden units and one output for the prediction
of the human breath rate time series. Linear and nonlinear specialized univariate methods are included.} 
\end{center}
\end{table}

\noindent Assuming that the data can be represented by a low--dimensional nonlinear map, a fact which is also
supported by evidence \cite{kantz}, an embedding dimension of $10$ is arbitrarily selected. An ANN with $10$ input
units, $20$ hidden neurons and one output unit is trained by SFP, HMCMC
and by the Broyden--Fletcher--Goldfarb--Shanno (BFGS) algorithms to approximate the nonlinear structure.
Two specialized methods for univariate time series are included for comparision: an Autoregressive (AR) 
model and a nonlinear predictor based on local approximations introduced by Kantz and Schreiber
\cite{kantz}, which we will denote here like Local Nonlinear Predictor (LNP).

\noindent In the Table \ref{table-breath} it has been considered the predicted Mean Squared Error (MSE), 
for the models in
which this statistic can be calculated: the AR, SFP and HMCMC. This quantity has been
estimated from the sample in the BFGS ANN model. For all the models, the out of sample MSE is
evaluated over the next $500$ time series values of the sample, 
except for the LNP, which do not needs a systematic parameter optimization
so the errors in the training set can be regarded as out of sample. The LNP method requires 
to set an embedding dimension, which
has been chosen with a value of ten, and a parameter that indicates the expected noise level. 
Following the results with the LNP reported by Kantz and 
Schreiber on this data set, a noise variance between $0.1$ and $0.5$ may be expected from data.
For the experiment reported in the Table \ref{table-breath}, a value of $0.5$ has been chosen, but no significant
difference in performance over the forementioned range has been observed.
For the AR model it has been used the Akaike information criterion in order to optimize the
model's complexity.

\noindent The parameters of HMCMC are taken from the regression example given in the documentation
of the software by it's author and are not intended to assure convergence but to give reasonable results
with short computation times.
The parameters for SFP were chosen in such a way that a comparably fast training 
for SFP-S should be expected,
using $L = 100$, $D = 0.005$ and $M = 100$. The computation times of both HMCMC and SFP turn out to be
in the order of $5$ minutes.

\noindent The BFGS ANN has been trained using the R package ``nnet'' with the default setting of a 
maximum of $100$ BFGS iterations for weight optimization. The resulting weights give an in sample MSE of
$0.04$, while the out of sample MSE is an order of magnitude above, indicating strong overfitting.
For SFP on the other hand, the predicted and observed MSE have the same order of magnitude, which is far
more satisfactory and is what it's expected from an adecuate Bayesian inference. From the results for the AR and the
LNP models it seems clear that the time series has indeed  nonlinear dependencies and these are captured
by SFP.

\noindent For this regression problem appears that the number of iterations for HMCMC were insufficient
to achieve an adequate convergence to the posterior, displaying a substantially inferior performance than
SFP in a similar computation time.

\subsection{Prediction of the output of a NMR--laser}

\noindent Other well known example of nonlinear data is given by the NMR--laser dataset 
\cite{kantz,badii}. 
The dataset consists on the signal produced by the output power of a nuclear magnetic 
resonance laser, which is modulated periodically. The signal is sampled $15$ times per period 
of modulation using a stroboscopic view. Here we have chosen the signal without noise
reduction.
In this 
example, there are strong arguments that indicate that the statistical properties of the
observed time series are mainly due to deterministic behavior, and that the noise component is 
rather low \cite{kantz}. In the experiments presented in Table \ref{table-laser}, there has been used a training set
of $200$ time series data points and test set based on $200$ time series points. Training and test sets are
completely disjoint. 
The selected ANN architecture in this case consists on two neurons for the
input layer, which corresponds to an embedding dimension with a size of two.
The hidden layer has been chosen with $20$ units and the output layer consists on a single neuron.
Like in the human breath rate example, the
BFGS ANN has been trained with the default setting of a 
maximum of $100$ BFGS iterations for weight optimization.
Following the discussion presented by Kantz and Schreiber \cite{kantz}, 
a noise variance of $0.1$ has been chosen 
for the LNP. For the AR model it has again been used the Akaike information criterion in order to optimize the
model's complexity.
The parameter values for SFP are $L= 100$, $M = 100$, $D = 0.005$ and the same setup for
HMCMC in the human breath rate signal example is used. The computation time for this example is around
$3$ minutes for HMCMC, $7$ minutes for SFP-S and SFP-MLE and $5$ minutes for SFP-I.
Again the conventionally trained ANN shows strong overfitting, displaying a performance
comparable with the one of the best linear model. The Bayesian ANN's instead
show an out of sample error that is essentially the same shown by 
the specialized nonlinear time series predictor. 

\begin{table}
\begin{center}

\begin{tabular}{|c|c|c|c|c|c|c|c|c|c|c|}  \hline
&	\textbf{In sample MSE}&	\textbf{Predicted MSE}&	\textbf{Out of sample MSE}		\\
	\hline 
\textbf{SFP-S} &	--    & 0.0591	&	0.0660	\\
	\hline
\textbf{SFP-MLE} &	--    & 0.0578	&	0.0490	\\
	\hline
\textbf{SFP-I} &	--    & 0.0600	&	0.0745	\\
	\hline
\textbf{HMCMC} &	--    & 0.0545	&	0.0668	\\
	\hline
\textbf{BFGS ANN}&	 0.0012   &	--   &	0.1873	\\
	\hline 
\textbf{AR(8)}&	 --   &	0.1622 &	0.1657	\\
	\hline 
\textbf{LNP}&	--    &	 --   &	0.0470	\\
\hline
\end{tabular} 
\vspace{10pt}
\caption{\label{table-laser} Performance of SFP and HMCMC Bayesian learning in comparision with a standard
training procedure in an ANN with $2$ inputs, $20$ hidden units and one output for the prediction
of the NMR--laser time series. Linear and nonlinear specialized univariate methods are included.} 
\end{center}
\end{table}    

\subsection{Behavior of large networks} \label{large}

\begin{figure}
	\centering
		\vskip 0.7cm
		\includegraphics[width=.7\textwidth]{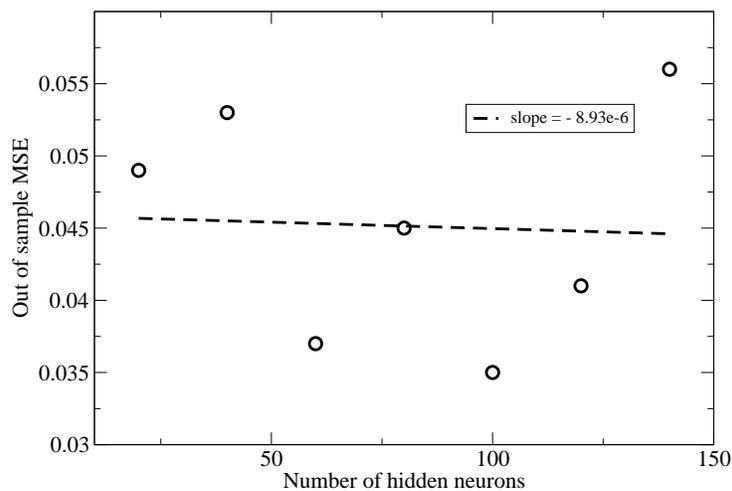}
	\caption{Plot of the observed out of sample MSE for SFP-MLE in the forecast of the NMR--laser 
time--series for increasing hidden layer size. 
The SFP parameters are selected in order to have a posterior density as sharp as possible.
The out of sample errors appear to be
independent of the neural network complexity.}
	\label{fig-laserSize}
\end{figure}

\noindent A major advantage of Bayesian inference is that in principle arbitrarily large models can be used without
the danger of ``overfitting''. 
Table \ref{table-laserSize} presents the best (in the sense of MLE Bayesian predicted squared error) ANN found after $20$ iterations of SFP
for the NMR--laser data, considering architectures with different sizes. 
An intensive SFP sampling is used, with $L = 300$ and $D = 1e-4$.
These
parameters are selected in order to have a posterior density as sharp as possible, which
is computationally demanding but useful to check how prone is SFP to overtraining.
Is clear that, in accordance with
it should be expected for a correct Bayesian estimation,
the overfitting effect is not present despite the
increasing model complexity.  
This claim is furtherly supported
in Fig. \ref{fig-laserSize}, where the out of sample errors for the different network sizes are plotted. 
A linear regression on the errors shows no evidence of an increment of the errors with 
the ANN size. A standard F-test for
this regression indicates that the null hypotesis of a constant slope can't be rejected at a $95 \%$ confidence.

\begin{table}
\begin{center}

\begin{tabular}{|c|c|c|c|c|c|c|c|c|c|c|}  \hline
\textbf{Number of hidden neurons}  & \textbf{Predicted MSE} & \textbf{Out of sample MSE}  \\
	\hline
\textbf{20}&	0.058    &	0.049 	\\
	\hline 
\textbf{40}&	 0.062   &	 0.053   	\\
	\hline 
\textbf{60}&	0.047   &     0.037   \\
	\hline 
\textbf{80}&	0.057    &  0.045	\\
	\hline 
\textbf{100}&	0.058    &  0.035   \\
	\hline 
\textbf{120}&	0.060   &    0.041	 \\
	\hline 
\textbf{140}&	0.064    &  0.056  \\
	\hline 
\end{tabular} 
\vspace{10pt}
\caption{\label{table-laserSize} The observed out of sample MSE for the SFP ANN in the forecast of the NMR--laser 
time--series for increasing hidden layer size.} 
\end{center}
\end{table}

\begin{figure}
	\centering
		\vskip 0.7cm
		\includegraphics[width=.7\textwidth]{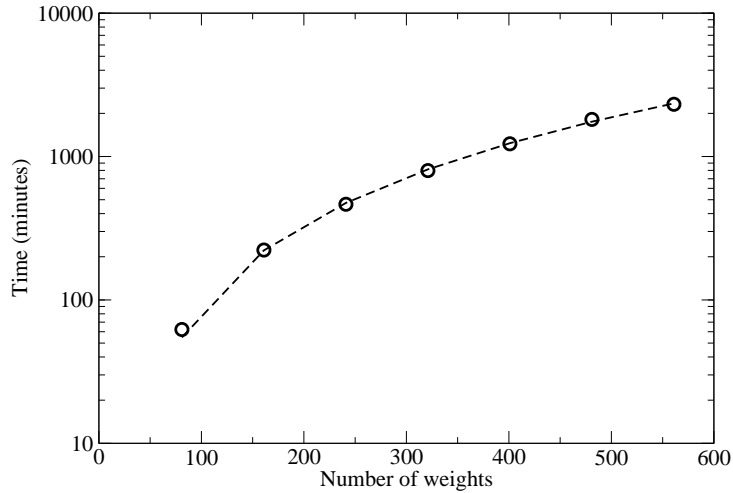}
	\caption{Dependence between computation time and ANN size for SFP-MLE training 
in the NMR--laser example. The SFP 
parameters are selected in order to have a posterior density as sharp as possible, which
is computationally demanding.
The dashed line represents the best quadratic fit, given by $Time = 
-26.57 + 0.46 N - 0.007 N^2$.
}
	\label{fig-laserTime}
\end{figure}

The number of loss function evaluations of a SFP sampling grows linearly 
with the potential's dimension \cite{berrones}. 
Therefore, Table \ref{table-laserSize} indicates that SFP is capable to perform a correct estimation of the posterior density 
with a total number of loss function evaluations that grows linearly with system's size.
The evaluations of the potential function
gives the major contribution to the computational cost.
In Fig. \ref{fig-laserTime} is presented the total computation time of each of the runs of Table \ref{table-laserSize}. 
The computation time grows slowly with the system's dimension, which
is consistent with the linear behavior predicted by SFP theory. This experimental result
is important because it shows the value of SFP sampling for large scale systems. In the context of global
optimization, it appears that SFP should be further adapted to alleviate 
at some extent the curse of dimensionality suffered by
any stochastic optimization method in order to be competitive with the current
best algorithms \cite{melchert}. However, Table \ref{table-laserSize} and Fig. \ref{fig-laserTime} suggest
that for density estimation purposes, the correct estimation of large dimensional densities via
SFP sampling is a polynomial time computational procedure, with a total number of loss function evaluations
that behave linearly. 
These properties may be valuable
in other applications besides Bayesian inference, like for instance Monte--Carlo simulations of large physical
systems.

\section{Discussion}

The framework for Bayesian learning based on SFP sampling introduced in this Letter is directly 
connected to equilibrium statistical mechanics. In particular,  
using a dimensionless Boltzmann constant $k=1$, it turns out that
an entropy $S$ can be introduced,
\begin{eqnarray}
S = - \int_{\vec{w}} p(\vec{w} | A) \ln p(\vec{w} | A)d\vec{w},
\end{eqnarray}
\noindent therefore obtaining the thermodynamic relation
\begin{eqnarray}
\left < V (\vec{w} | A) \right> = S - \left < \ln p(A | \vec{w}) \right>.
\end{eqnarray}
\noindent The exploitation of the link of SFP Bayesian learning with equilibrium statistical mechanics appears to
be promising taking into account that SFP provides analytic expressions for the marginals, from which mean--field
approximations to the quantities of interest may be derived.

\noindent An additional interesting
research question is the study of more general forms of uncertainty affecting the stochastic search 
in the weight space. 
In this regard, it should be noticed
that the SFP formalism is in principle not limited to the estimation of the stationary density of a
diffusion on a potential under white Gaussian additive noise. Through the use of generalizations
to the Fokker--Planck equation based
on the expansion of the master equation, like for instance
the Van Kampen expansion \cite{vankampen}, several other stochastic search processes may be 
considered. If the posterior densities resulting from such generalized processes
are more adequate in some situations seems to be an appealing question to further study.

\noindent A techical issue in which there may be room for the improvement of the SFP approach
regards the selection of the most appropriate basis function family for the approximation of the
conditionals. In this Letter it has been used the Fourier basis only because of its simplicity, but
in principle any other basis can be used.

\noindent
Other relevant research line that follows from the results presented so far consists on the 
application of SFP learning to large
and complex systems. If the observed polynomial behavior of computation time holds in general, it 
would be valuable to apply the SFP technique on very large inference models, taking advantage
of the intrinsic parallel nature of the SFP sampling algorithm.

\section*{Acknowledgement} 

This work was partially supported by the National Council of Science and Technology of
Mexico and by the UANL--PAICYT program.


\end{document}